\documentclass[11pt,a4paper]{article}

\usepackage[utf8]{inputenc}
\usepackage[T1]{fontenc}
\usepackage{lmodern}
\usepackage[margin=2.5cm]{geometry}
\usepackage{microtype}
\usepackage{amsmath,amssymb}
\usepackage{graphicx}
\usepackage{booktabs}
\usepackage{tabularx}
\usepackage{subcaption}
\usepackage{tikz}
\usepackage{pgfplots}
\pgfplotsset{compat=1.18}
\usepackage[numbers,sort&compress]{natbib}
\usepackage{setspace}
\usepackage[font=small,labelfont=bf,labelsep=period]{caption}
\usepackage{titlesec}
\usepackage[hidelinks]{hyperref}
\titleformat{\section}{\large\bfseries}{\thesection.}{0.5em}{}
\titleformat{\subsection}{\normalsize\bfseries}{\thesubsection.}{0.5em}{}
\titleformat{\subsubsection}{\normalsize\itshape}{\thesubsubsection.}{0.5em}{}

\newcommand{\articletype}[1]{}
\newcommand{\orcid}[1]{}
\newcommand{\affilbox}{}
\newcommand{\affil}[1]{\gappto\affilbox{\par #1}}
\usepackage{etoolbox}
\newcommand{\emailaddr}{}
\newcommand{\email}[1]{\renewcommand{\emailaddr}{#1}}
\newcommand{\keywordlist}{}
\newcommand{\keywords}[1]{\renewcommand{\keywordlist}{#1}}
\newcommand{\ack}[1]{\section*{Acknowledgments}#1}
\newcommand{\funding}[1]{\section*{Funding}#1}
\newcommand{\roles}[1]{\section*{Author contributions}#1}
\newcommand{\data}[1]{\section*{Data availability statement}#1}
 
\renewcommand{\title}[1]{\gdef\thetitle{#1}}
\renewcommand{\author}[1]{\gdef\theauthors{#1}}
\newenvironment{articleabstract}{%
  \begin{center}
  {\LARGE\bfseries\thetitle\par}
  \end{center}\vspace{0.6em}
  \begin{flushleft}
  {\large\theauthors\par}\vspace{0.8em}
  {\small\itshape\affilbox\par}\vspace{0.4em}
  {\small E-mail: \emailaddr\par}
  \end{flushleft}\vspace{0.8em}
  \noindent\textbf{Abstract.}\ }{%
  \par\vspace{0.6em}\noindent\textbf{Keywords:} \keywordlist\par\vspace{1.2em}}

\begin{document}

\articletype{Paper}

\title{From prompts to physical laws: exploring mechanics through AI-generated videos}

\author{Laura B Alvarado-Cruz$^{1,2}$\orcid{0000-0001-5565-1001}, Francisco M Mu\~noz-P\'erez$^{1}$\orcid{0000-0002-0313-5029}, Josep Ll Su\~ner$^{3}$\orcid{0009-0003-8678-1361}, Pedro Yuste$^{4}$\orcid{0000-0002-6688-9426}, Juan C Castro-Palacio$^{1}$\orcid{0000-0002-0132-9989} and Juan A Monsoriu$^{1,*}$\orcid{0000-0003-3350-7951}}

\affil{$^1$Centro de Tecnolog\'ias F\'isicas, Universitat Polit\`ecnica de Val\`encia, 46022 Valencia, Spain}

\affil{$^2$Institut Universitari de Ci\`encia dels Materials (ICMUV), Universitat de Val\`encia, 46010 Valencia, Spain}

\affil{$^3$Instituto de Ingenier\'ia Mec\'anica y Biomec\'anica, Universitat Polit\`ecnica de Val\`encia, 46022 Valencia, Spain}

\affil{$^4$Instituto Universitario de Investigaci\'on de Tecnolog\'ias de la Informaci\'on y Comunicaciones, Universitat Polit\`ecnica de Val\`encia, 46022 Valencia, Spain}

\affil{$^*$Author to whom any correspondence should be addressed.}

\email{\href{mailto:jmonsori@fis.upv.es}{jmonsori@fis.upv.es}}


\begin{articleabstract}
This proof-of-concept study examines whether motion generated by text-to-video artificial intelligence (AI) can be analysed quantitatively using the tools of an introductory physics laboratory. Three videos, generated with PixVerse.ai and Pippit, depict a wooden block sliding with friction, a sphere falling into glycerin and a descending skydiver. Positions were tracked with Tracker and fitted to equations of motion derived from Newton’s second law, including gravity and buoyancy where relevant. For each video, the best-fitting model reproduces the tracked positions with a root-mean-square error of 5 mm or less, yet independent consistency checks, under the stated assumptions, reveal clear departures from the requested scenarios: the wooden block slides almost without friction, with an estimated friction coefficient about two orders of magnitude below typical values for wooden surfaces; the sphere behaves as if it were nearly neutrally buoyant; and the skydiver, filmed as if by a camera falling alongside, does not fall through the frame, so the effect of gravity cannot be observed. Model comparison also indicates that the block and skydiver videos do not clearly distinguish between different resistive force models. These results suggest that good agreement with an analytical model is not sufficient to ensure physical consistency. We propose the workflow as an educational activity in which students critically analyse AI-generated motion using physical models and quantitative consistency tests; its effect on learning remains to be evaluated in the classroom.
\end{articleabstract}

\section{Introduction}\label{sec1}

In the last decade, the integration of AI in physics education has transformed from a futuristic promise to an increasingly important driver of pedagogical innovation \cite{Chen, krstic2022}. Several studies report that the strategic use of AI can improve students' critical skills, academic performance, and depth of knowledge, while optimizing the overall teaching and learning experience \cite{Arini, Muhammad}. By providing adaptive learning environments and personalized resources, AI allows students to interact with complex concepts in a more intuitive and efficient manner, effectively bridging the gap between analytical abstraction and practical application \cite{zawacki, Holmes}.

In the teaching of introductory mechanics, one of the greatest challenges remains the transition from ideal, frictionless models to a comprehensive understanding of real-world phenomena, such as dissipative forces \cite{robertson2002force, MIT}. Traditionally, this process has relied on physical laboratories, which can be complemented by other resources to enrich learning and foster a deeper understanding of these phenomena \cite{trumper, hofstein, Jong}. Recent studies have shown that generative AI can support education through the creation of customised visual resources that help students explain and understand complex technical concepts \cite{Nieto}. In parallel, generative AI has also been employed to develop interactive experimental resources and virtual laboratory environments for physics education through natural-language prompting \cite{suner}. Building on these emerging approaches, generative AI for video, represented by platforms such as PixVerse.ai \cite{pixverse2026} and Pippit \cite{pippit2026}, offers new opportunities to create dynamic and measurable learning scenarios. These tools could provide synthetic material on which students practise data analysis without the logistical barriers of a conventional laboratory setting, provided that the generated motion is itself examined critically. Video analysis with Tracker is a well-established tool in physics teaching \cite{brown2009, wee2012}, and computer-generated animations have already been used as ``artificial video'' for activities in which real footage is difficult to obtain \cite{gallis2010}. The present work extends this idea to text-to-video generative AI, with an essential difference: the generated motion is not explicitly computed from a physical model.%

At the same time, recent benchmarks have shown that state-of-the-art video generation models frequently violate basic physical laws, and that visual realism is not a reliable indicator of physical correctness \cite{kang2025, motamed2026, bansal2025, meng2025}. These models are trained primarily to produce plausible-looking images, not to integrate equations of motion. Any educational use of AI-generated videos as a source of ``experimental'' data must therefore include a critical step in which the generated motion is tested against independent physical constraints.

This study explores whether AI-generated videos can serve as material for quantitative analysis in which students test the generated motion against physical models, rather than use it to confirm physical laws. To illustrate this approach,
 the present work addresses three fundamental braking force models that often present conceptual difficulties for students: constant friction, drag proportional to velocity, and drag proportional to the square of the velocity. Each regime has a characteristic signature in the trajectory, which makes it possible to ask whether a generated video displays the behaviour that its prompt implies.%

The central objective is to provide a framework where the student, using video tracking tools, can extract kinematic data from the resources generated by PixVerse.ai and Pippit to verify whether these synthetic experiments align with the expected analytical behaviour. A key element of the framework is that each model is written in its complete form, including all the forces acting on the body, so that it yields not only a fitting function but also independent consistency relations (for instance, between the fitted parameters and the gravitational acceleration or the density ratio) that distinguish a physically consistent trajectory from one that merely has the right functional shape. By extending previous work on artificial video for video analysis \cite{gallis2010} and on AI-generated physics laboratories \cite{suner} to text-to-video generative models, this article shows how AI acts not only as a visual aid but as a source of data whose output must itself be scrutinized. The proposed methodology integrates AI-generated content, motion tracking, and quantitative model evaluation into a coherent learning workflow. The work is presented as a proof of concept: it establishes the technical feasibility of the workflow and identifies the physical checks that it requires, while its effect on student learning is left for a future classroom evaluation.

\section{Methodology}\label{sec2}

\begin{figure}[htbp]
    \centering
    \includegraphics[width=1\linewidth]{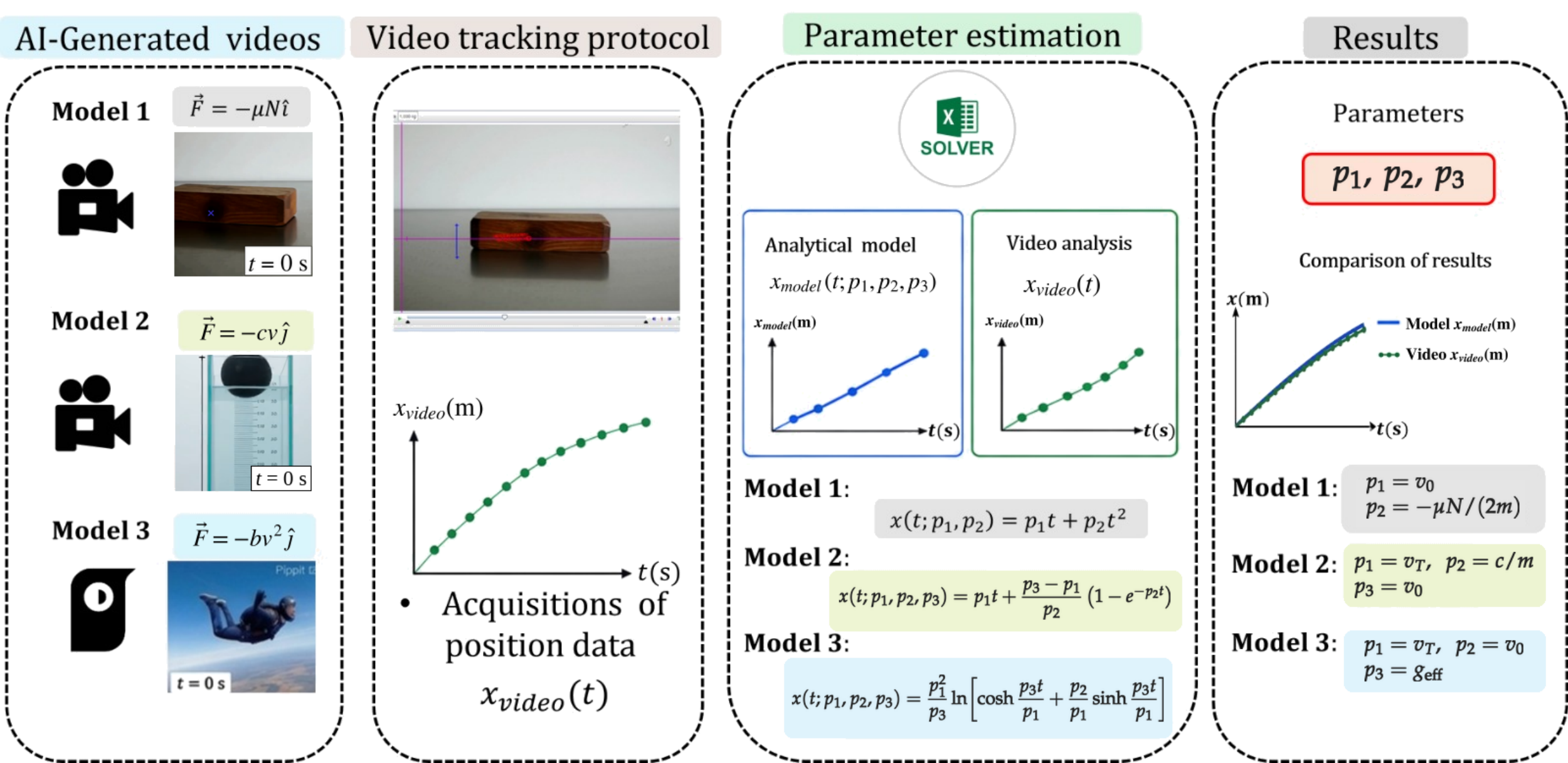}
    \caption{Workflow of the proposed methodology, in four phases: AI video creation, video tracking and data extraction, parameter estimation through curve fitting and, finally, physical interpretation of the parameters and consistency tests. The fitting functions and the meaning of the parameters correspond to the complete models of Section~\ref{sec2} (Eqs.~\ref{EQ1p}, \ref{eq:m2_fit} and~\ref{eq:m3_fit}). The resistive forces oppose the motion and act along the horizontal axis ($\hat{\imath}$) for the block and along the vertical axis ($\hat{\jmath}$, positive downward) for the sphere and the skydiver; in each case, $x$ denotes the position along that axis.}
    \label{fig:1}
\end{figure}

The proposed workflow is organized into four sequential phases: (i) generation of synthetic videos using generative AI, (ii) extraction of kinematic data through video analysis with the Tracker software, (iii) estimation of model parameters by means of non-linear curve fitting, and (iv) physical interpretation of the recovered parameters and consistency tests. A schematic overview of the workflow is presented
in Figure~\ref{fig:1}.
\subsection{AI-Generated videos}\label{sec:videos}

To generate the three scenarios, generative artificial intelligence platforms were employed using a set of keywords associated with different dynamical behaviours. These prompts specified the visual and physical characteristics of each scenario while avoiding explicit implementation of the underlying equations of motion, as summarized in Table~\ref{tab:prompts}.

\begin{table}[ht!]
\centering
\caption{Prompts used to generate the AI-generated scenarios analysed in this study.}
\label{tab:prompts}
\begin{tabularx}{\textwidth}{@{}l X p{2.5cm}@{}}
\toprule
\textbf{Physical Model} & \textbf{Prompt Used} & \multicolumn{1}{l}{\textbf{AI Platform Used}}\\ \midrule
Model 1: Constant Friction & \textit{``Create a $20 \times 10 \times 5$ cm wooden block that slides horizontally on a smooth or waxed flat surface. The view of the wooden block should be from the front.''} & PixVerse.ai \cite{pixverse2026}\\ \addlinespace[0.5em]
Model 2: Linear Drag & \textit{``A silicone rubber sphere falls into a transparent glass
container filled with a viscous, high-viscosity glycerin
solution. The diameter of the sphere is $5$~\text{cm}.''} & PixVerse.ai \cite{pixverse2026}\\ \addlinespace[0.5em]
Model 3: Quadratic Drag & \textit{``Long, slow-motion video of a skydiver
falling at high speed through the atmosphere,
with an approximate body mass of $80$ \text{kg} including equipment.
The perspective is from a lateral view.''} & Pippit \cite{pippit2026}\\ \bottomrule
\end{tabularx}
\end{table}

\begin{figure}[htbp]
    \centering
    \begin{subfigure}{\textwidth}
        \centering
        \textbf{\small (a) Model 1: Constant Friction (Sliding Block)}\par\vspace{1.5pt}
        \includegraphics[width=0.9\linewidth, height=3.8cm, keepaspectratio]{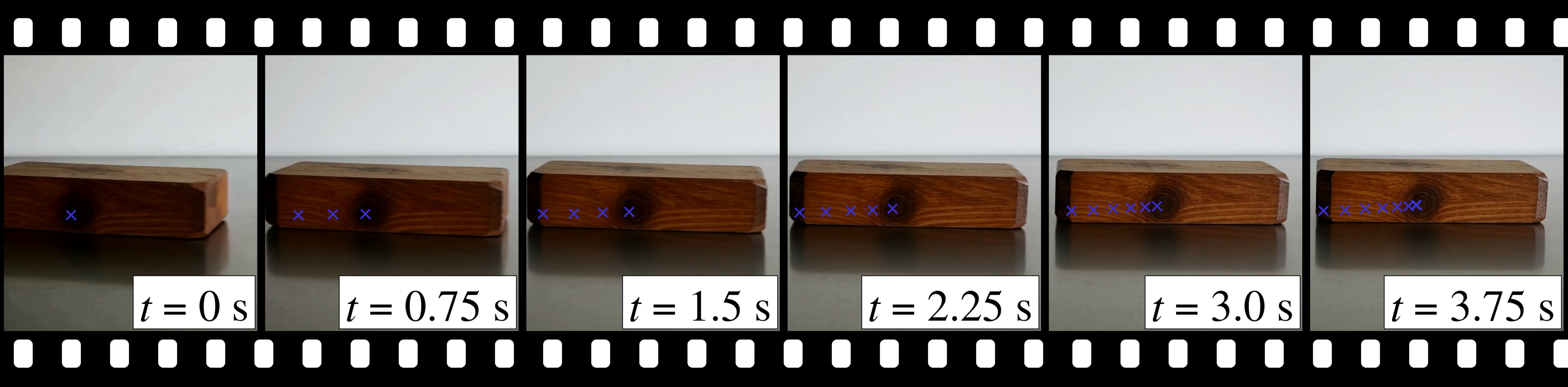}
    \end{subfigure}
    \vspace{10pt}
    \begin{subfigure}{\textwidth}
        \centering
        \vspace{10pt}
        \textbf{\small (b) Model 2: Linear Drag (Sphere falling in viscous glycerin)}\par\vspace{1.5pt}
        \includegraphics[width=0.9\linewidth, height=4.2cm, keepaspectratio]{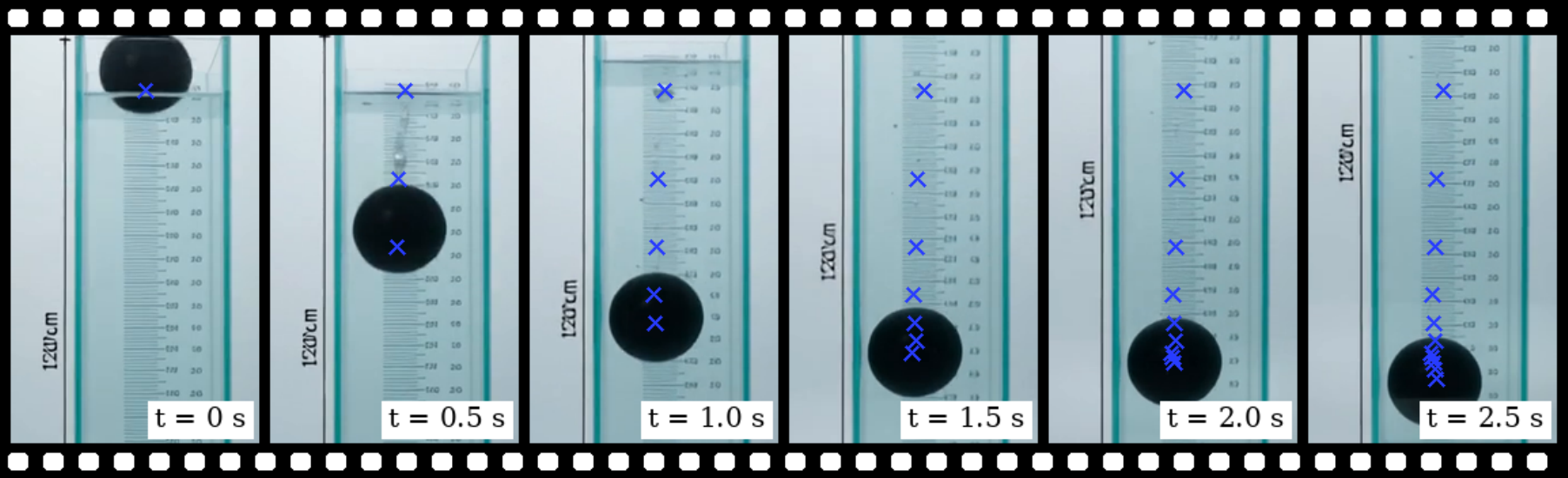}
    \end{subfigure}
    \vspace{10pt}
    \begin{subfigure}{\textwidth}
        \centering
        \textbf{\small (c) Model 3: Quadratic Drag (Skydiver)}\par\vspace{2pt}
        \includegraphics[width=0.9\linewidth, height=3.8cm, keepaspectratio]{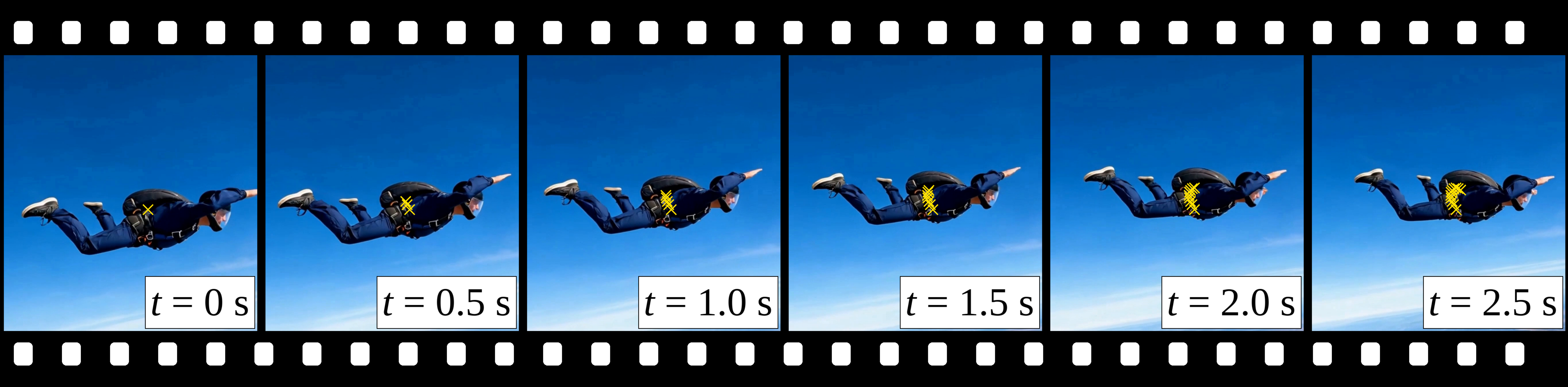}
    \end{subfigure}
    \caption{Representative filmstrips from the AI-generated videos for the three scenarios; each panel shows its time. The blue ``x'' markers show the tracked positions every 0.25~s up to that time: in (a), those of the dark knot on the front face of the block (each panel is centred on the block, which comes to rest at $t\approx3.4$~s) (Video~1); in (b), those of the sphere, measured from its position at $t=0$ with the 5~cm diameter as scale (Video~2); in (c), those of the skydiver's backpack, drawn in yellow for contrast with the sky (Video~3).}
    \label{fig:2}
\end{figure}

For each prompt, between three and seven video generations were performed. The first output in which the target object remained continuously visible throughout the entire motion and could be reliably tracked in Tracker was retained for analysis. The selection of the videos was based exclusively on object visibility and trackability, without considering their physical realism, the quality of the subsequent fits, or their agreement with the expected dynamical behaviour. No generation seed was set or recorded. The objective of this procedure was not to identify the most physically realistic output, but rather to obtain trajectories suitable for quantitative analysis using video-tracking techniques. This allowed the extracted position data to be compared with the corresponding analytical models and enabled an assessment of whether and to what extent AI-generated motion reproduced the expected physical behaviour. In addition, the resulting datasets provide students with opportunities to practise data analysis, parameter estimation and model validation, which may foster a deeper understanding of the underlying laws of motion. The videos were generated using the web interfaces of PixVerse V6 \cite{pixverse2026} (Models 1 and 2) and Pippit \cite{pippit2026} (Model 3) as available on 13 May 2026. At that date, Pippit employed the ByteDance Seedance video model, prior to the release of Seedance 2.5. Output resolutions were $1280\times720$ pixels for Model 1, $640\times360$ pixels for Model 2 and $1080\times1482$ pixels (portrait format) for Model 3. Video durations were 5 s for Models 1 and 2 and 3.4 s for Model 3. No frame interpolation or post-processing was applied. The three original AI-generated videos are provided as supplementary material (Videos 1--3), and representative frames are shown in Figure~\ref{fig:2}.

\subsection{Video tracking protocol}
The kinematic analysis of the synthetic videos generated by the AI platforms was performed using Tracker, an open-source
video analysis and modelling tool \cite{Tracker}. The software interface provides an integrated environment where the
visual scene is synchronized with real time data plotting and numerical recording. To transform the AI-generated pixels into physical observables, we implemented a three-step protocol that enables the extraction of quantitative information from the generated videos:

\begin{itemize}
    \item Calibration and Scaling: Upon importing the synthetic video, spatial dimensions were established using the
Calibration Stick tool. A nominal reference length such as the length or height of the tracked object was used to define the metric scale of the virtual environment. This step is fundamental to ensure that all positions and coordinates are recorded in SI (International System of Units). Because the reference lengths are nominal values rather than measured ones, this calibration assumes that the generator has rendered the objects at those sizes; any error in this assumption propagates to all lengths, velocities and accelerations. The time axis is taken from the native frame rate of each video (24~frames/s for Models~1 and~2, 30~frames/s for Model~3), which in a generated video need not correspond to real time, particularly when slow motion is requested. The calibration references were the 20~cm length of the block (Model~1) and the 5~cm diameter of the sphere (Model~2); for the skydiver (Model~3), since no object of known size appears in the scene, a nominal body length of 1.80~m, from heel to helmet, was used. A $\pm10\%$ error in the reference length changes all lengths, velocities and accelerations by the same fraction.
\item Coordinate System Alignment: A Cartesian reference frame was overlaid on the scene. The origin was anchored at the position of the tracked point in the first frame ($t=0$), and the $x$ axis was aligned with the trajectory: horizontal for the sliding block (Model~1) and vertical, positive downwards, for the falling sphere and the skydiver (Models~2 and~3). This configuration simplifies the analysis by describing the motion in terms of a single spatial coordinate, which
is essential for obtaining clean $x(t)$ datasets for subsequent analysis.

\item Point-Mass Tracking: The object's trajectory was captured by defining a ``Point Mass'' on the moving body. By tracking a fixed point on the object frame-by-frame, the software simultaneously populated a coordinate table and generated kinematic plots. The scatter in the fit residuals is of the order of a few millimetres (Section~\ref{sec:results}); tracking errors may increase in frames where the generated object changes shape or appearance. For the block, the dark knot near the centre of its front face provided a natural high-contrast feature to track. Tracking was performed with the Tracker autotracker; for the skydiver, the backpack was used as the tracked feature.

\end{itemize}

The output of this protocol provided the empirical basis for the non-linear regression and the physical interpretation stage conducted in the following phase of the study.

\subsection{Analytical Models and Parameter Estimation}
The motion of an object is governed by Newton's second law, $\sum\vec{F} = m\vec{a}$, where $\sum\vec{F}$ is the sum of the external forces acting on it, such as weight, buoyancy or a resistive force. The form of these forces determines the complexity of the resulting equation of motion \cite{Halliday2014, Young2016, serway}. Three distinct
force models are considered in this work, each leading to
a different analytical expression for the position as a
function of time $x(t)$ and providing the basis for the subsequent evaluation of the AI-generated motion. To facilitate the non-linear
optimization process, the physical parameters are
reparametrized as $p_1$, $p_2$ and, for Models~2 and~3, $p_3$, as summarized for
each model below. To minimize the discrepancy between
the video-extracted positions and the corresponding analytical predictions, the Microsoft Excel \textit{Solver} add-in is employed
using the GRG Nonlinear (Generalized Reduced Gradient) algorithm \cite{glantz},
minimizing the sum of squared residuals:
\begin{equation}
    S = \sum_{q=1}^{n}\left(x_{video,q} -
    x_{model,q}\right)^2,
    \label{EQS}
\end{equation}
where $x_{video,q}$ is the position extracted from
the video at time step $q$, $x_{model,q}$ is the
value predicted by the analytical model, and $n$ is
the total number of frames included in the fit. This quantity $S$
is minimised by adjusting the fitting parameters
for each of the three models described below, thereby enabling direct comparison between the AI-generated trajectories and the theoretical predictions.
All fits were carried out in Excel. Because $R^2$ has limitations as a goodness-of-fit measure for non-linear least squares, it is reported only because it is the measure most familiar to students; the root-mean-square error (RMSE), the residuals and their lag-1 autocorrelation, and the Akaike information criterion (AIC), $\mathrm{AIC}=n\ln(S/n)+2k$, where $k$ is the number of fitted parameters \cite{akaike1974}, are the quantities used to judge each fit. To check that the fitted functional forms are specific to each scenario, six candidate models (constant velocity, constant friction, drag-only linear and quadratic drag, and the complete linear- and quadratic-drag models) were fitted to every video. Solver was set to minimise $S$ with the GRG Nonlinear algorithm, central derivatives and a convergence tolerance of $10^{-10}$, with the non-negativity constraint on the variables disabled, and was run repeatedly until $S$ stopped decreasing; where the result depended on the starting point, the Multistart option was used, with bounds on the variables ($g_{\rm eff}>0$ in the quadratic-drag models). The initial values were order-of-magnitude guesses used only to start the optimisation: $v_0=1$~m/s together with $\mu=0.1$ (constant friction), $c/m=1$~s$^{-1}$ (linear drag only), $b/m=10^{-3}$~m$^{-1}$ (quadratic drag only), $v_T=0.1$~m/s and $c/m=1$~s$^{-1}$ (complete linear drag), and $v_T=5$~m/s and $g_{\rm eff}=9.81$~m/s$^2$ (complete quadratic drag). Standard errors were obtained in the same spreadsheet from the Jacobian $J$ of the model, computed by finite differences, as $\mathrm{cov}(\mathbf{p})=(J^{\mathrm T}J)^{-1}S/(n-k)$; for Eq.~\ref{EQ1p} this coincides with \texttt{LINEST}. Because consecutive tracking errors are correlated, these standard errors were multiplied by $\sqrt{(1+\phi)/(1-\phi)}$, where $\phi$ is the lag-1 autocorrelation coefficient of the residuals (a first-order autoregressive approximation to the reduced number of independent points), and the 95\% confidence intervals were built from the corrected errors with Student's $t$ distribution for $n-k$ degrees of freedom. This correction is only approximate for non-linear fits, so the intervals should be read as indicative. Likewise, the AIC assumes independent errors; with correlated residuals the $\Delta$AIC values are used to rank the models, not as a significance test.

\subsubsection{Model 1: Constant Friction}
When a solid object slides on a horizontal surface, the
dominant resistive mechanism is kinetic friction, described
by a constant force $\vec{F} = -\mu N \hat{\imath}$, where
$\mu$ is the kinetic friction coefficient and $N$ is the
normal force \cite{Halliday2014, serway}. Applying
Newton's second law, the block moves with constant acceleration $a=-\mu N/m$; on a horizontal surface $N=mg$, so $a=-\mu g$. Integrating once gives the velocity, $v(t)=v_0-(\mu N/m)\,t$, and integrating again with $x(0)=0$ gives the position:
\begin{equation}
    x(t) = v_0\,t - \frac{\mu N}{2m}\,t^2,
    \label{EQ1}
\end{equation}
where $v_0$ is the initial velocity and $m$ is the
mass of the object. This parabolic profile is
characteristic of motion under constant deceleration and provides a straightforward benchmark for evaluating the physical consistency of the generated trajectory.
For the fitting procedure, Eq.~\ref{EQ1} is
reparametrized as:
\begin{equation}
    x(t; p_1, p_2) = p_1\,t + p_2\,t^2,
    \label{EQ1p}
\end{equation}
where $p_1 = v_0$ is the initial velocity and
$p_2 = -\mu N/(2m)$ represents half the constant
acceleration. Equation~\ref{EQ1} is valid only while the block is moving, $t\leq t_s$, where $t_s=v_0/(\mu g)$ is the time at which the block stops after travelling the stopping distance $d_s=v_0^2/(2\mu g)$: once it stops, kinetic friction ceases and the parabola must not be extrapolated. Because Eq.~\ref{EQ1p} is linear in $p_1$ and $p_2$, it can be fitted by ordinary least squares (e.g.~the \texttt{LINEST} function in Excel, with the constant term set to zero), which also returns the standard errors of the parameters.
For a block on a horizontal surface $N=mg$, so the deceleration is $\mu g$, independent of the mass. The model therefore carries a direct consistency check: once $v_0$ has been measured, a kinetic friction coefficient $\mu$ fixes the stopping time $t_s=v_0/(\mu g)$ and the stopping distance
\begin{equation}
  d_s = \frac{v_0^2}{2\mu g},
  \label{eq:m1_stop}
\end{equation}
which can be compared directly with the video, both for the value of $\mu$ obtained from the fit and for the values tabulated for the materials named in the prompt.

\subsubsection{Model 2: Linear Drag}
We consider a sphere of mass $m$, volume $V$ and density $\rho_s = m/V$ descending through a liquid of density $\rho_f$. The vertical axis is taken positive downwards, with position $x(t)$ and velocity $v=\dot{x}$. For sufficiently slow motion in a viscous liquid the resistive force is proportional to the velocity, $F_r=-cv$ with $c>0$ \cite{stokes1851, landau}. While the sphere descends ($v\geq 0$) it is acted upon by its weight $mg$, the Archimedes buoyancy force $\rho_f V g$ and the drag force, so that Newton's Second Law gives
\begin{equation}
  m\frac{dv}{dt} = (m-\rho_f V)\,g - c\,v .
  \label{eq:m2_ode}
\end{equation}
Defining the characteristic time $\tau$ and the terminal velocity $v_T$,
\begin{equation}
  \tau = \frac{m}{c}, \qquad
  v_T = \frac{(m-\rho_f V)\,g}{c} = \frac{(\rho_s-\rho_f)\,V g}{c},
  \label{eq:m2_tau_vT}
\end{equation}
Eq.~\ref{eq:m2_ode} becomes $\dot v + v/\tau = v_T/\tau$. Multiplying by the integrating factor $e^{t/\tau}$ and imposing $v(0)=v_0$ gives $v(t)=v_T+(v_0-v_T)\,e^{-t/\tau}$, and a further integration with $x(0)=0$ yields
\begin{equation}
  x(t) = v_T\,t + (v_0-v_T)\,\tau\left(1-e^{-t/\tau}\right).
  \label{eq:m2_x}
\end{equation}
The velocity relaxes exponentially, with time constant $\tau$, from its initial value $v_0$ to the terminal value $v_T$; at long times the position grows linearly, $x\simeq v_T t + (v_0-v_T)\tau$. Two limits are worth noting. If the sphere is released from rest ($v_0=0$), $x(t)=v_T\left[t-\tau\left(1-e^{-t/\tau}\right)\right]$. If instead the sphere is neutrally buoyant ($\rho_s=\rho_f$, hence $v_T=0$), Eq.~\ref{eq:m2_x} reduces to $x(t)=v_0\tau\left(1-e^{-t/\tau}\right)$, i.e.\ the sphere simply coasts to rest over an asymptotic distance $v_0\tau$. Because the scene described in the prompt (Table~\ref{tab:prompts}) shows a sphere \emph{falling into} the liquid, a non-zero initial velocity must be allowed for, and the fitting function is
\begin{equation}
  x(t;p_1,p_2,p_3) = p_1\,t + \frac{p_3-p_1}{p_2}\left(1-e^{-p_2 t}\right),
  \label{eq:m2_fit}
\end{equation}
with $p_1=v_T$ (m/s), $p_2=c/m=1/\tau$ (s$^{-1}$) and $p_3=v_0$ (m/s).

Equations~\ref{eq:m2_ode}--\ref{eq:m2_tau_vT} imply a relation between the fitted parameters that contains no additional adjustable quantity,
\begin{equation}
  p_1\,p_2 = \frac{v_T}{\tau} = g\left(1-\frac{\rho_f}{\rho_s}\right),
  \label{eq:m2_consistency}
\end{equation}
which provides a direct test of the physical consistency of the generated motion: the product of two fitted parameters should agree with the value determined by the gravitational acceleration and the density ratio expected for the materials named in the prompt. For a sphere of radius $r$ in the Stokes regime, $c=6\pi\eta r$, where $\eta$ is the dynamic viscosity of the liquid \cite{stokes1851, landau}, and therefore
\begin{equation}
  \eta = \frac{p_2\, m}{6\pi r} = \frac{2}{9}\,\rho_s\, r^2\, p_2 .
  \label{eq:m2_eta}
\end{equation}
These expressions rest on three assumptions that students should check explicitly: (i) the Reynolds number $\mathrm{Re}=\rho_f v d/\eta$, with $d=2r$ the diameter of the sphere, is small ($\mathrm{Re}<1$), otherwise the drag is no longer linear in $v$ \cite{munson}; (ii) the added mass of the displaced fluid, $C_A\rho_f V$ with $C_A=1/2$ for a sphere \cite{landau}, is negligible---if not, $m$ must be replaced by $m+C_A\rho_f V$ in $\tau$ (but not in $v_T$); and (iii) the container walls and bottom are far from the sphere, since wall proximity increases the drag \cite{happel1983} and contact with the bottom ends the validity of Eq.~\ref{eq:m2_x}. Accordingly, only the frames recorded before the sphere reaches the bottom are included in the fit.

\subsubsection{Model 3: Quadratic Drag}
At high Reynolds numbers the aerodynamic drag is proportional to the square of the velocity, $F_r=-b\,v|v|$, with $b=\tfrac12\rho_{\rm air}C_D A$, where $C_D$ is the drag coefficient and $A$ the projected area \cite{munson, anderson}. For a body of mass $m$ falling vertically (axis positive downwards, $v\geq0$), assuming a constant drag coefficient and projected area, and neglecting air buoyancy ($\rho_{\rm air}/\rho_{\rm body}\sim10^{-3}$, where $\rho_{\rm air}$ and $\rho_{\rm body}$ are the densities of the air and of the body), Newton's Second Law reads
\begin{equation}
  m\frac{dv}{dt} = m g - b\,v^2 ,
  \label{eq:m3_ode}
\end{equation}
which defines the terminal velocity
\begin{equation}
  v_T = \sqrt{\frac{m g}{b}} .
  \label{eq:m3_vT}
\end{equation}
For $v_0<v_T$ the solution is $v(t)=v_T\tanh\!\left(gt/v_T+\operatorname{artanh}(v_0/v_T)\right)$ and for $v_0>v_T$ it is $v(t)=v_T\coth\!\left(gt/v_T+\operatorname{arcoth}(v_0/v_T)\right)$. For $v_0=v_T$, the velocity remains constant. In all three cases integration with $x(0)=0$ leads to the single expression
\begin{equation}
  x(t) = \frac{v_T^2}{g}\,
  \ln\!\left[\cosh\!\left(\frac{g t}{v_T}\right)
  + \frac{v_0}{v_T}\,\sinh\!\left(\frac{g t}{v_T}\right)\right],
  \label{eq:m3_x}
\end{equation}
which reduces to $x(t)=(v_T^2/g)\ln\cosh(gt/v_T)$ for a release from rest. Whatever the value of $v_0$, the velocity tends to $v_T$, so at long times the position grows linearly, $x\simeq v_T t + \text{const}$. The horizontal-motion expression $x(t)=(m/b)\ln(1+b v_0 t/m)$, in which the velocity decays to zero, is recovered from Eq.~\ref{eq:m3_x} only in the limit $g\to0$ at fixed $b/m$, i.e.\ in the absence of gravity; it is therefore not applicable to a falling body.

For fitting, we use
\begin{equation}
  x(t;p_1,p_2,p_3) = \frac{p_1^2}{p_3}\,
  \ln\!\left[\cosh\!\left(\frac{p_3 t}{p_1}\right)
  + \frac{p_2}{p_1}\,\sinh\!\left(\frac{p_3 t}{p_1}\right)\right],
  \label{eq:m3_fit}
\end{equation}
with $p_1=v_T$ (m/s), $p_2=v_0$ (m/s) and $p_3=g_{\rm eff}$ (m/s$^2$). Equations~\ref{eq:m3_ode}--\ref{eq:m3_fit} assume downward motion ($v\geq0$). If the body moves upward, the drag acts downward and Eq.~\ref{eq:m3_ode} becomes $m\,dv/dt=mg+bv^2$ until the velocity changes sign; the fits use the solution with the drag sign that corresponds to the direction of motion. The gravitational acceleration is deliberately left as a free parameter: since $g$ is known independently, the condition
\begin{equation}
  g_{\rm eff} = g = 9.81~\text{m/s}^2
  \label{eq:m3_consistency}
\end{equation}
is a parameter-free test of the physical consistency of the generated motion. It also probes the time and length calibrations of the video, because an unknown playback slow-down factor $s$ or a scale error $\lambda$ in the spatial calibration would give $g_{\rm eff}=\lambda g/s^2$. Only if Eq.~\ref{eq:m3_consistency} is satisfied is it meaningful to fix $p_3=g$ and to derive $b=mg/v_T^2$ and $C_D=2mg/(\rho_{\rm air}A\,v_T^2)$. In the opposite limit, $v_T\to\infty$ (negligible drag), Eq.~\ref{eq:m3_x} reduces to uniformly accelerated motion, $x=v_0t+g\,t^2/2$. Finally, the test assumes that positions are measured from a camera at rest with respect to the ground; if the camera moves with the body, only the relative acceleration is measured, which for a fall at terminal velocity is close to zero even in a real recording.

These three models represent a natural pedagogical
progression in complexity: from constant deceleration
(Model 1) to exponential relaxation towards a terminal velocity (Model 2) and hyperbolic relaxation towards a terminal velocity (Model 3), covering the resistive force regimes
most commonly encountered in introductory
physics courses \cite{Halliday2014, Young2016, serway}. In each case, the complete equation of motion provides, in addition to the fitting function, a consistency relation (Eqs.~\ref{eq:m1_stop}, \ref{eq:m2_consistency} and \ref{eq:m3_consistency}) that does not depend on the goodness of fit and that also helps to judge whether a given video contains enough information to apply the model.

\section{Results and discussion}\label{sec:results}

\begin{figure}[htbp]
    \centering
    \includegraphics[width=0.60\linewidth]{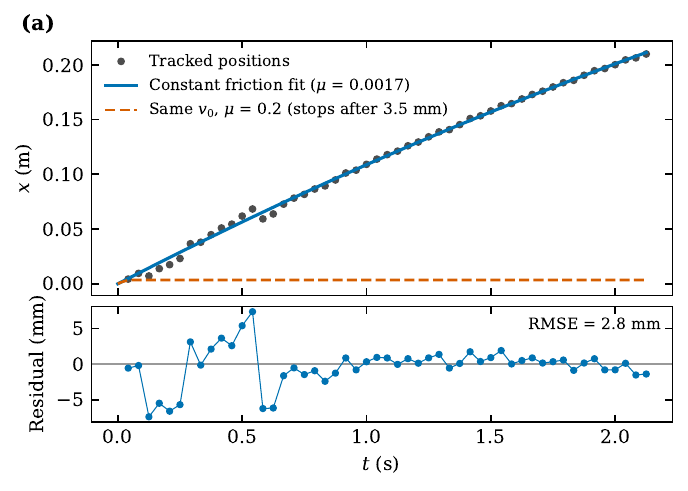}\\[2pt]
    \includegraphics[width=0.60\linewidth]{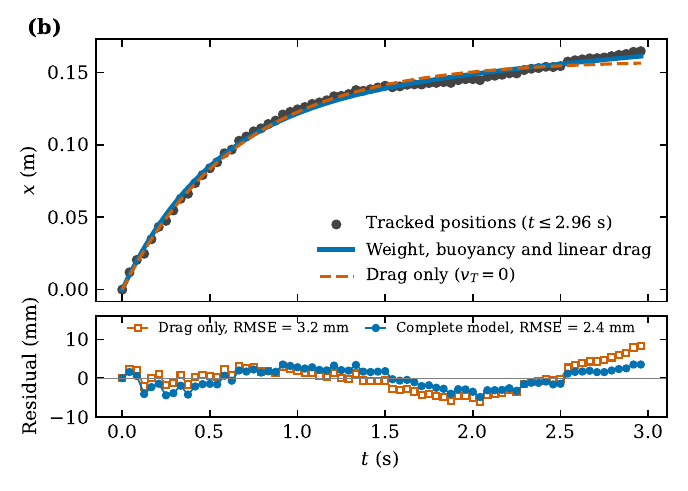}\\[2pt]
    \includegraphics[width=0.60\linewidth]{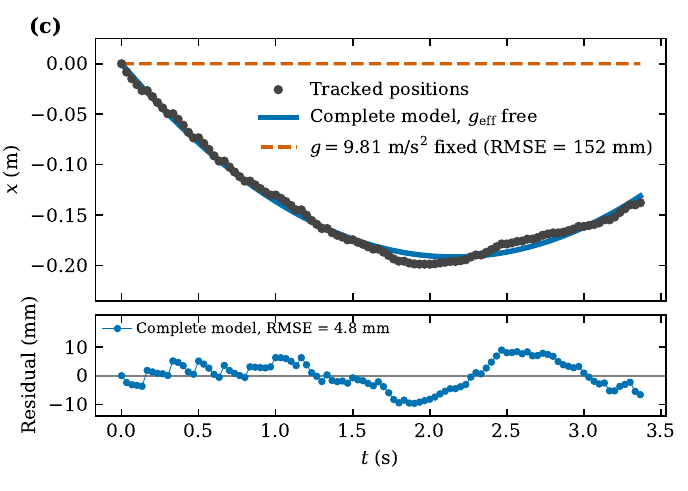}
    \caption{Comparison between the video-extracted trajectories and the analytical models; the lower panel of each plot shows the fit residuals. (a) Model~1: sliding block (dots) and fit of Eq.~\ref{EQ1p} (solid line), giving $\mu\approx0.0017$. The dashed line is the trajectory predicted by Eq.~\ref{EQ1} for the same initial velocity and a conservative wood-on-wood coefficient $\mu=0.2$, for which the block stops after 3.5~mm. (b) Model~2: sphere descending in the glycerin. Dots: the $n=72$ frames recorded before the sphere reaches the bottom of the container ($t\leq2.96$~s). Solid line: Eq.~\ref{eq:m2_fit} (weight, buoyancy and linear drag); dashed line: drag-only limit $v_T=0$ fitted to the same interval. (c) Model~3: vertical position of the skydiver relative to the frame (positive downward). Solid line: complete quadratic-drag model with $g_{\rm eff}$ free, which converges to the limit $v_T\to\infty$, i.e.\ to constant acceleration, with $g_{\rm eff}\approx0.082$~m/s$^2$; dashed line: Eq.~\ref{eq:m3_x} with $g=9.81$~m/s$^2$ fixed, which cannot follow the upward drift. The lower panel of (b) shows the residuals of both models.}
    \label{fig:3}
\end{figure}

\begin{figure}[htbp]
    \centering
    \includegraphics[width=0.60\linewidth]{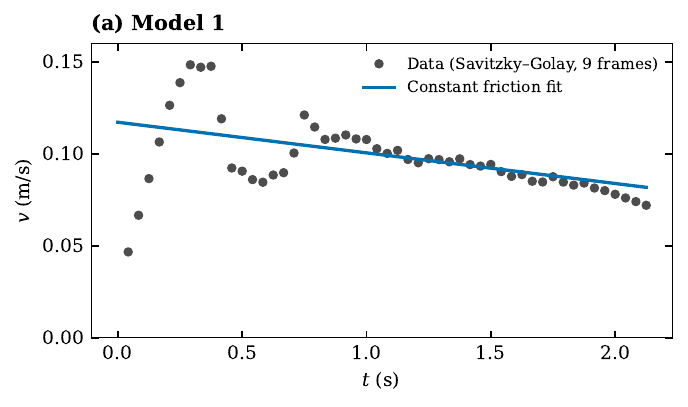}\\[2pt]
    \includegraphics[width=0.60\linewidth]{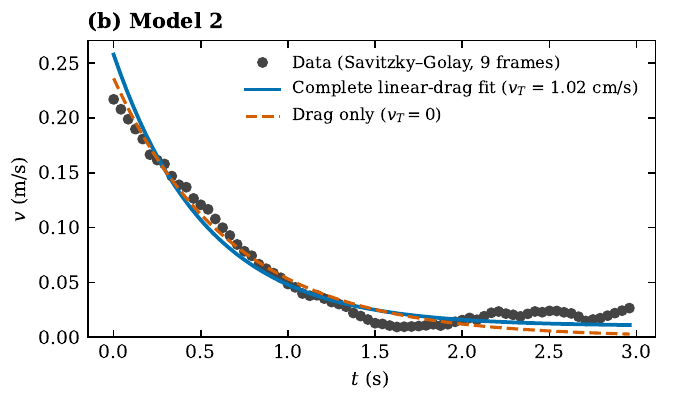}\\[2pt]
    \includegraphics[width=0.60\linewidth]{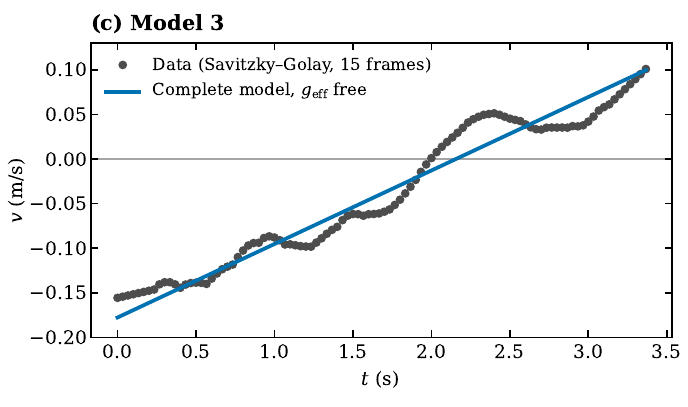}
    \caption{Velocity obtained by numerical differentiation of the tracked positions (dots; Savitzky--Golay filter) compared with the velocity predicted by the fitted models, $v=p_1+2p_2t$ for Models~1 and~3 and $v=p_1+(p_3-p_1)e^{-p_2t}$ for Model~2. (a) Model~1: after irregular tracked positions during the first 0.75~s, which account for much of the apparent curvature in Figure~\ref{fig:3}(a), the velocity decreases slowly. (b) Model~2: the velocity relaxes to a minimum of $\approx0.9$~cm/s at $t\approx1.6$~s and then increases slightly, which Eq.~\ref{eq:m2_x} cannot reproduce; frames after the bottom contact are not included. (c) Model~3: the velocity relative to the frame increases almost linearly, with oscillations, from $\approx-0.16$~m/s (upward) to $\approx+0.10$~m/s; the solid line corresponds to $g_{\rm eff}\approx0.082$~m/s$^2$.}
    \label{fig:4}
\end{figure}

Least-squares regression was applied to the $x(t)$ data extracted with \textit{Tracker}, and each fit was judged by its RMSE, its residuals and the AIC (Section~\ref{sec2}); the coefficient of determination ($R^2$) is quoted for reference \cite{glantz, miles}. Figures~\ref{fig:3}(a)--(c) compare the tracked data with the optimised analytical curves for the three dynamical regimes. In every case $R^2>0.99$ is obtained for the model that best describes the data. However, as shown below, a high $R^2$ only indicates that the model closely reproduces the tracked positions. Whether the AI-generated motion is physically consistent with the scenario requested in the prompt is assessed through the values of the fitted parameters and the consistency relations derived in Section~\ref{sec2}.
Table~\ref{tab:models} compares the six candidate models for every video, and Figure~\ref{fig:4} shows the velocity obtained by numerical differentiation of the tracked positions (Savitzky--Golay filter, second-order polynomial, 9 frames for Models~1 and~2 and 15 frames for Model~3), which allows the approach to a terminal velocity to be examined directly.%

\begin{table}[htbp]
\centering
\caption{Comparison of the six candidate models fitted to each video: root-mean-square error (RMSE, mm) and difference in Akaike information criterion with respect to the best model of each video ($\Delta$AIC; models with differences below $\approx2$ have comparable empirical support). The model selected for further discussion is shown in bold.}
\label{tab:models}
\begin{tabular}{@{}lrrrrrr@{}}
\toprule
 & \multicolumn{2}{c}{A: Model 1} & \multicolumn{2}{c}{B: Model 2} & \multicolumn{2}{c}{C: Model 3}\\
\cmidrule(lr){2-3}\cmidrule(lr){4-5}\cmidrule(l){6-7}
Model & RMSE & $\Delta$AIC & RMSE & $\Delta$AIC & RMSE & $\Delta$AIC\\
\midrule
Constant velocity                    & 5.1 & 60.7 & 35.7 & 385.1 & 52.8 & 487.9\\
Constant friction, Eq.~\ref{EQ1p}$^{e}$    & \textbf{2.8} & \textbf{0.0} & 12.3 & 233.5 & \textbf{4.8} & \textbf{0.0}\\
Linear drag only ($v_T=0$)           & 2.8 & 1.0 & 3.2 & 40.8 & 17.7 & 266.5\\
Quadratic drag only ($g=0$)          & 2.8 & 2.0 & 5.7 & 122.7 & 23.9 & 327.9\\
Linear drag, complete, Eq.~\ref{eq:m2_fit}    & 2.8$^{a}$ & 2.1$^{a}$ & \textbf{2.4} & \textbf{0.0} & 4.8$^{b}$ & 2.0$^{b}$\\
Quadratic drag, complete, Eq.~\ref{eq:m3_fit} & 2.8$^{a}$ & 4.2$^{a}$ & 5.7 & 124.7 & 4.8$^{b}$ & 2.0$^{b}$\\
\midrule
Frames $n$; frame interval (s)       & \multicolumn{2}{r}{51; 1/24} & \multicolumn{2}{r}{72; 1/24} & \multicolumn{2}{r}{102; 1/30}\\
Fitted interval (s)                  & \multicolumn{2}{r}{0.04--2.13} & \multicolumn{2}{r}{0--2.96$^{c}$} & \multicolumn{2}{r}{0--3.37}\\
Lag-1 residual autocorrelation$^{d}$ & \multicolumn{2}{r}{0.47} & \multicolumn{2}{r}{0.87} & \multicolumn{2}{r}{0.94}\\
Parameter correlation, $\mathrm{corr}(p_1,p_2)^{d}$ & \multicolumn{2}{r}{$-0.97$} & \multicolumn{2}{r}{$0.90$} & \multicolumn{2}{r}{$-0.97$}\\
\bottomrule
\end{tabular}\\[2pt]
\begin{minipage}{\textwidth}\footnotesize
$^{a}$With physically admissible parameters ($1/\tau>0$, $g_{\rm eff}>0$) Solver converges to the constant-friction or the drag-only limit of the model, so the extra parameter brings no improvement. $^{b}$With $g_{\rm eff}$ free, both complete models converge to the constant-acceleration limit ($c/m\to0$, $v_T\to\infty$), so the extra parameter brings no improvement. $^{c}$Frames before the bottom contact. $^{d}$For the model in bold; obtained from the covariance matrix of the fit, it is not the correlation between $x$ and $t$. $^{e}$For Model~3 this row is motion with constant acceleration, $p_2=g_{\rm eff}/2$.
\end{minipage}
\end{table}

\subsection{Parameter interpretation}
\label{sec:params}

\subsubsection*{Model 1: Constant Friction}
Figure~\ref{fig:3}(a) shows the position of the sliding block (Video~1). For a block sliding on a horizontal surface, $N = mg$, and therefore $p_2 = -\mu g/2$. Fitting Eq.~\ref{EQ1p} with positions in metres yields $p_1=v_0=0.1171\pm0.0013$~m/s and $p_2=-0.0083\pm0.0008$~m/s$^2$ (nominal standard errors; $R^2=0.9980$, RMSE $=2.8$~mm). The trajectory is therefore almost uniform, and part of its curvature comes from irregularities of the tracked position during the first 0.75~s (residuals of up to $\pm7$~mm), discussed below. The friction coefficient recovered from the fit is
\begin{equation}
    \mu = \frac{2\,|p_2|}{g} = \frac{2 \times 0.0083}{9.81}
    \approx 0.0017,
    \label{eq:m1_mu}
\end{equation}
with a 95\% confidence interval of $[0.0012,\,0.0022]$. This value is almost two orders of magnitude below the kinetic coefficients tabulated in introductory textbooks for wood (e.g.~$\mu_k\approx0.1$ for waxed wood on wet snow and $\mu_k\approx0.3$ for wood on wood; table~6.1 of \cite{openstax}). Since the prompt does not specify the material of the surface, it is worth noting that even the lowest kinetic coefficients in the same table, $0.02$--$0.04$ for steel on ice, ice on ice or Teflon on steel, are more than ten times larger. The inconsistency is clearest when Eq.~\ref{eq:m1_stop} is applied. With the measured $v_0$ and $\mu=0.2$, a conservative value below the tabulated 0.3, the block should stop within $t_s\approx0.06$~s after travelling only $d_s\approx3.5$~mm (dashed line in Figure~\ref{fig:3}(a)). In the video, however, it covers $\approx0.21$~m in 2.1~s and comes to rest only at $t\approx3.4$~s, after $\approx0.27$~m, still inside the frame; with the fitted $\mu=0.0017$ it would need $\approx7$~s and $\approx0.4$~m to stop, so the deceleration of the generated motion does not remain constant beyond the fitted interval. Within the fitted interval, the generated motion is therefore compatible with a nearly constant deceleration, as Eq.~\ref{EQ1} assumes, but this deceleration corresponds to an almost frictionless surface rather than to a wooden block on a smooth or waxed table.

The deceleration is statistically significant. Ordinary least squares, which for Eq.~\ref{EQ1p} is equivalent to \texttt{LINEST}, returns the same parameters and standard errors, with a Student test statistic $t_{\rm S}=-10.9$ for $p_2$, and a nested $F$-test rejects the constant-velocity model ($F=119$; $\Delta\mathrm{AIC}=61$, Table~\ref{tab:models}). The nominal $p$-values of these tests ($p<10^{-13}$) assume independent residuals and are therefore too small, since the lag-1 autocorrelation of the residuals is 0.47; with the standard error corrected for this correlation, $t_{\rm S}=-6.6$ and the confidence interval for $\mu$ still excludes zero. The two parameters are strongly anticorrelated ($\mathrm{corr}(p_1,p_2)=-0.97$). Since $t_s\approx7$~s exceeds the fitted interval of 2.1~s, Eq.~\ref{EQ1} is valid throughout the fitted interval. However, the linear and quadratic drag-only models describe the same data comparably well ($\Delta\mathrm{AIC}\leq2.0$), so this video shows that the block decelerates but does not allow the resistive force law to be identified. The velocity obtained from the data (Figure~\ref{fig:4}(a)) shows that much of the apparent curvature comes from the first 0.75~s. In this interval the tracked position moves backwards once ($t=0.125$~s) and jumps by 13~mm between two consecutive frames ($t\approx0.27$~s), so that the apparent speed rises from $\approx0.06$ to $\approx0.15$~m/s before falling again. In the video the block appears to move smoothly during this interval, so these irregularities most likely come from the autotracker rather than from the generated motion. After 0.75~s the motion is regular and the velocity decreases slowly from $\approx0.11$ to $\approx0.07$~m/s. Fitting this interval alone ($n=34$, with the initial position as an additional free parameter) gives $\mu=0.0027$ with a 95\% confidence interval of $[0.0023,\,0.0030]$ and RMSE $=0.7$~mm, so the conclusion that the surface is almost frictionless does not depend on the first frames. A $\pm10\%$ calibration error changes $\mu$ by $\pm10\%$, which does not affect these conclusions.

\subsubsection*{Model 2: Linear Drag}
Figure~\ref{fig:3}(b) shows the sphere (Video~2), entering the liquid at $t=0$ (Figure~\ref{fig:2}(b)), decelerating in the glycerin until it touches the bottom of the container at $t\approx 3.0$~s. At that point the trajectory shows a maximum followed by a small rebound and a plateau at $x\approx 0.16$~m. The fit was therefore restricted to the $n=72$ frames with $t\leq 2.96$~s. The released-from-rest form ($p_3=0$) fails to describe the data ($R^2=0.22$), which confirms that the sphere is still moving with a substantial velocity at the start of the tracked interval, as expected after falling into the liquid. The three-parameter function, Eq.~\ref{eq:m2_fit}, yields $p_1=v_T=1.02$~cm/s, $p_2=1/\tau=1.88$~s$^{-1}$ and $p_3=v_0=0.258$~m/s, with 95\% confidence intervals of $[0.10,\,1.93]$~cm/s, $[1.39,\,2.37]$~s$^{-1}$ and $[0.224,\,0.291]$~m/s, respectively ($R^2=0.9965$, RMSE $=2.4$~mm). Compared with the drag-only limit $x=v_0\tau(1-e^{-t/\tau})$ fitted to the same interval, the full model reduces the RMSE from 3.2 to 2.4~mm and is strongly preferred by the Akaike criterion ($\Delta\mathrm{AIC}\approx40$). The estimate of $v_T$ is, however, sensitive to the end of the fitting window: it moves from $-2.3$ to $+1.0$~cm/s as the cut-off is varied between 2.0 and 2.96~s. The robust conclusion is therefore that the terminal velocity is small, of the order of $1$~cm/s, compared with the magnitude of the initial velocity.

The consistency relation, Eq.~\ref{eq:m2_consistency}, gives $p_1p_2=0.019$~m/s$^2$, i.e.\ $\rho_f/\rho_s\approx 0.998$. Given the sensitivity of $v_T$ to the fitting window (see above, $p_1p_2$ between $-0.028$ and $+0.019$~m/s$^2$) and the confidence interval of the final fit, the data allow $\rho_f/\rho_s$ anywhere between about 0.996 and 1.003. If the added mass is included, the right-hand side of Eq.~\ref{eq:m2_consistency} becomes $g(\rho_s-\rho_f)/(\rho_s+C_A\rho_f)$, which changes these ratios by less than 0.3\%. In either case the generated motion corresponds to a sphere that is almost neutrally buoyant in the liquid. This is not the behaviour expected from the nominal densities. A silicone rubber sphere with $\rho_s\approx 1200$~kg/m$^3$ \cite{wacker} in an 85~wt\% glycerin solution ($\rho_f\approx 1222$~kg/m$^3$ at 20~$^\circ$C \cite{bosart1927}) would give $p_1p_2=-0.18$~m/s$^2$, i.e.\ the sphere would rise rather than sink. Since the density of commercial silicone rubbers spans roughly 1050--1600~kg/m$^3$ \cite{wacker}, the prompt does not fix $\rho_s$, and the video should be read as implying $\rho_s\approx\rho_f$ rather than as reproducing the nominal materials.

Applying Eq.~\ref{eq:m2_eta} with $r=0.025$~m and, consistently with the result above, $\rho_s\approx\rho_f\approx1220$~kg/m$^3$ ($m\approx0.080$~kg) gives an apparent viscosity $\eta\approx0.32$~Pa$\cdot$s, or $\approx0.48$~Pa$\cdot$s if the added mass is included. This number is an effective coefficient for this scene, not a measurement of the liquid's viscosity: the initial Reynolds number is $\mathrm{Re}_0=\rho_f v_0 d/\eta\approx 30$--$50$, well outside the Stokes regime, and the sphere is close to the container walls. Moreover, tabulated glycerin--water viscosities range from $\approx0.11$~Pa$\cdot$s for 85~wt\% at 20~$^\circ$C to $1.41$~Pa$\cdot$s for pure glycerol \cite{segur1951}, so a value of this order would match some mixture whatever its origin.

The model comparison and the velocity extracted from the data support these conclusions. Among the six candidate models, the complete linear-drag model is clearly preferred (Table~\ref{tab:models}: $\Delta\mathrm{AIC}=40$ with respect to the drag-only limit and $>120$ with respect to quadratic drag), and the velocity (Figure~\ref{fig:4}(b)) relaxes from $\approx0.25$~m/s to a small value; the frames after the bottom contact are excluded from the analysis. The relaxation is not monotonic, however: the velocity reaches a minimum of $\approx0.9$~cm/s at $t\approx1.6$~s and then increases to $\approx2.7$~cm/s before the bottom contact, a behaviour that Eq.~\ref{eq:m2_x} cannot produce. This explains why the estimate of $v_T$ depends on the fitting window and why the residuals remain strongly autocorrelated (lag-1 coefficient 0.87): even the best model does not capture the motion completely, and $p_1$ and $p_2$ are correlated ($\mathrm{corr}(p_1,p_2)=0.90$). The viscosity estimate is not robust to the corrections that the scene requires. The sphere occupies a fraction $d/D\approx0.65$ of the container width $D$ (estimated from Figure~\ref{fig:2}(b); the prompt does not specify the container, so its size can only be estimated from the image), for which the low-Reynolds-number wall correction of Francis increases the drag by a factor $K\approx15$ \cite{francis1933, clift1978}; a finite-Reynolds-number correction of the Schiller--Naumann type, $1+0.15\,\mathrm{Re}^{0.687}$ \cite{clift1978}, increases it further by a factor of about 3 at the speeds of the video ($\mathrm{Re}\approx30$--$50$). Since these corrections depend on the container size and the Reynolds number, and their combination is not a validated model, they cannot be used to obtain a corrected value of $\eta$. Moreover, since $\eta\propto r^2$ and $r$ is taken from the prompt, a $\pm10\%$ error in the sphere diameter alone changes $\eta$ by $+21\%$ or $-19\%$. The video therefore does not constrain the viscosity of the liquid.

\subsubsection*{Model 3: Quadratic Drag}
Figure~\ref{fig:3}(c) shows the vertical position of the skydiver (Video~3) relative to the frame. The skydiver does not fall through the frame: the tracked point rises by $\approx0.20$~m during the first 2.0~s and then descends by $\approx0.06$~m, while drifting horizontally by up to $\approx0.15$~m. When the gravitational acceleration is fixed at $g=9.81$~m/s$^2$, Eq.~\ref{eq:m3_x} cannot reproduce this behaviour (RMSE $=152$~mm, $R^2<0$): with gravity, a body moving upward decelerates at a rate of at least $g$ and then falls, whereas the tracked point decelerates about a hundred times more slowly. When $g_{\rm eff}$ is left free, the complete quadratic-drag model converges to the limit $v_T\to\infty$, i.e.\ to motion with constant acceleration, $x=v_0t+g_{\rm eff}t^2/2$, with $v_0=-0.178$~m/s (upward; 95\% interval $[-0.189,\,-0.167]$~m/s) and $g_{\rm eff}=0.082$~m/s$^2$ (95\% interval $[0.074,\,0.091]$~m/s$^2$), about 120 times smaller than $g$ ($R^2=0.9918$, RMSE $=4.8$~mm). The quadratic-drag parameter is therefore not determined by the data. The residuals oscillate with an amplitude of $\approx1$~cm and are strongly autocorrelated (lag-1 coefficient 0.94), reflecting a swaying of the body that none of the models of Section~\ref{sec2} includes. In other words, the consistency condition, Eq.~\ref{eq:m3_consistency}, is not satisfied by the tracked motion.

Several features of the generated scene may contribute to this result. The skydiver remains in the lower part of the frame throughout the video, as if filmed by a camera falling alongside, and no stationary object is visible, so the motion cannot be referred to a fixed frame. For a camera and a skydiver that both fall at terminal velocity, the relative acceleration is zero, so a small $g_{\rm eff}$ is also what a real recording of this kind would show: Eq.~\ref{eq:m3_consistency} assumes a camera at rest and cannot be applied to motion measured relative to a moving camera. In addition, the requested slow motion makes the time scale of the video unknown; explaining the result by time scaling alone would require a slow-down factor $s=\sqrt{g/g_{\rm eff}}\approx11$. The motion of the camera, the unknown time scale and a generated motion that does not follow gravity may each contribute to the small value of $g_{\rm eff}$, and their relative weight is difficult to assess from this video alone. In any case, the video does not show the fall requested in the prompt, and its motion cannot be used to test the effect of gravity or to estimate $b$ or $C_D$.

The model comparison supports this interpretation (Table~\ref{tab:models}). Constant acceleration is the preferred model; the complete linear- and quadratic-drag models reduce to it, and their $\Delta\mathrm{AIC}=2.0$ reflects only the penalty for the additional parameter. The drag-only models, which cannot reverse the direction of motion, and the constant-velocity model are strongly rejected ($\Delta\mathrm{AIC}\geq266$). The skydiver video therefore does not single out the quadratic drag law implied by the scenario. A $\pm10\%$ error in the spatial calibration changes $v_0$ and $g_{\rm eff}$ by the same fraction and does not affect these conclusions.

The results across the three models lead to a consistent picture. In all cases an analytical model describes the tracked trajectory with $R^2>0.99$, and the generated motion is compatible with a simple resistive regime: a nearly constant deceleration for the block, exponential relaxation for the sphere and a slow, almost uniformly accelerated drift relative to the frame for the skydiver. The physical parameters and the consistency relations, however, reveal clear departures from the scenarios requested in the prompts: an almost frictionless wooden block ($\mu\approx0.002$), a sphere that is nearly neutrally buoyant in glycerin, and a skydiver who does not fall through the frame, so that the effect of gravity cannot be observed. This agrees with recent benchmarks reporting that generative video models reproduce the appearance of physical events without reliably obeying the underlying laws \cite{kang2025, motamed2026}. From an educational perspective, this is what could make the workflow valuable. In it, students must write the complete equation of motion rather than a simplified formula; they must identify the range of validity of the model (bottom contact, Reynolds number, calibration of time and length, reference frame); and they must judge, on quantitative grounds, whether a realistic-looking video obeys the laws of physics. These are the same critical skills required when analysing real experimental data, and the workflow can be implemented with widely available tools even where access to physical laboratory resources is limited.

Model comparison adds an important nuance to this picture. Only one of the three videos singles out the intended velocity dependence of the resistive force: the sphere video is best described by linear drag, whereas the block video is compatible with constant friction and with both velocity-dependent drag laws, and the skydiver video is best described by constant acceleration, with no identifiable drag law. Thus, a good fit with the intended model indicates that the data are consistent with its mathematical form; to identify the underlying force law with greater confidence, it is advisable to check that the alternative models do not describe the data equally well.

\subsection{Limitations}
\label{sec:limitations}
This study is a proof of concept with some limitations. First, a single video was analysed for each scenario, and only two platforms were used (PixVerse for Models 1 and 2, Pippit for Model 3); platform, prompt and scenario are therefore partly confounded, and no conclusion can be drawn about the variability between generations or about differences between platforms. Second, proprietary generators are stochastic and change over time, so the same prompt will not reproduce the same video; the three original videos are therefore provided as supplementary material (Videos 1--3) to enable readers to repeat the video analysis and compare their results with those presented in this article. Third, the spatial calibration relies on dimensions specified in the prompt, which the generator is not guaranteed to honour, and the time scale is uncertain when slow motion is requested; when the camera moves with the object, as in the skydiver video, the motion can only be measured relative to the frame. Fourth, the fit residuals provide only an indirect indication of tracking uncertainty, without repeated independent tracking. Finally, this work focuses on the development and validation of the workflow; its classroom implementation and the evaluation of its impact on learning, using a validated instrument and a comparison group, are the next step in this line of work.

\section{Conclusions}\label{sec:conclusions}

The present work has explored the viability of integrating generative AI with video analysis and numerical optimization tools in the context of introductory mechanics courses. Across the three scenarios studied, least-squares fitting reproduced the tracked trajectories with small residual errors and close agreement with the tracked positions. However, when the complete equations of motion are used, the recovered parameters reveal clear departures from the scenarios described in the prompts: the apparent friction coefficient of the wooden block is approximately $0.002$, the falling sphere behaves as if it were almost neutrally buoyant, and the skydiver, filmed as if by a camera falling alongside, does not fall through the frame, so the effect of gravity cannot be observed. Agreement with an analytical model therefore indicates, at most, that a generated trajectory is compatible with its mathematical form; physical consistency must be established through independent tests based on the underlying physics, such as the stopping distance, the density ratio $\rho_f/\rho_s = 1 - p_1p_2/g$, or the condition $g_{\mathrm{eff}} = g$, together with explicit checks of the model's range of validity.

The prompt acts as an implicit specification of the physical system, and crafting it is a form of experimental design that may encourage students to think about the parameters defining a scenario before performing any measurements. Whether more detailed prompts lead to more physically consistent videos is an interesting question for future work, which could generate several videos for each prompt at different levels of detail and compare their physical consistency using the relations proposed here.

Neither of the platforms used, PixVerse.ai and Pippit, is designed or advertised as a physics simulator. They are general-purpose video generators whose training data and objectives are not publicly available. The analysis therefore focuses on evaluating the resulting motion, without addressing the internal mechanisms that produce it. 

From this perspective, the generated videos can be described by simple analytical laws, although the consistency tests suggest that this agreement should be interpreted with caution: the generated motion may follow the functional form of a model without necessarily reproducing its underlying physics. This distinction between mathematical agreement and physical consistency is one of the central ideas of the proposed workflow. Detecting such discrepancies is itself a potential learning objective.

The integration of these platforms with Tracker and the Excel Solver add-in offers students in introductory physics courses a structured workflow involving mathematical modelling, quantitative analysis, and critical evaluation of results, applicable in both face-to-face and remote learning contexts. By combining AI-generated scenarios, motion tracking, and parameter estimation, the proposed methodology provides a framework through which students can practise model-based evaluation of data as a complement to activities involving experimental setups.

By designing prompts, extracting motion data, and testing physical models against consistency relations, students are encouraged to connect observation, measurement, and physical interpretation. In this way, AI-generated videos become objects of scientific inquiry rather than passive visual resources. In future work, it would be valuable to explore in the classroom how this workflow contributes to these skills, for example through a study with participants, a validated instrument and a comparison group. As generative AI becomes increasingly common in educational settings, the ability to evaluate its outputs using established physical principles may become an important component of physics education.

\ack{L B Alvarado-Cruz acknowledges the support of the Consejo de Ciencia, Tecnolog\'ia e Innovaci\'on de Hidalgo (CITNOVA), Mexico. The authors thank the Institute of Education Sciences, Universitat Polit\`ecnica de Val\`encia, Spain, for the support of the Educational Innovation and Quality Team SmartSTEM.}

\section*{AI disclosure statement}
Generative AI tools (PixVerse V6 and Pippit) were used to generate the videos analysed in this study, as described in Section~\ref{sec:videos}. The generated videos were quantitatively analysed and critically evaluated by the authors. ChatGPT was also used to assist with language editing and manuscript revision. The authors take full responsibility for the scientific content of the manuscript.

\funding{This research was funded by the Ministerio de Ciencia e Innovaci\'on, Spain (grant PID2022-142407NB-I00), the Generalitat Valenciana, Spain (grant CIPROM/2022/30) and the Universitat Polit\`ecnica de Val\`encia, Spain (grant PIME/24-25/423).}

\roles{L B Alvarado-Cruz (L.A.C.), F M Mu\~noz-P\'erez (F.M.M.), J Ll Su\~ner (J.Ll.S.), P Yuste (P.Y.), J C Castro-Palacio (J.C.C.P.) and J A Monsoriu (J.A.M.). Conceptualization, L.A.C. and F.M.M.; methodology, L.A.C., J.A.M., J.Ll.S., P.Y., J.C.C.P. and F.M.M.; software, L.A.C.; validation, L.A.C.; formal analysis, L.A.C.; investigation, L.A.C.; data curation, L.A.C.; writing---original draft, L.A.C.; writing---review and editing, L.A.C., J.A.M., J.Ll.S., P.Y., J.C.C.P. and F.M.M.; visualization, L.A.C. and F.M.M.; supervision, J.A.M., J.Ll.S., P.Y. and J.C.C.P. All authors have read and agreed to the submitted version of the manuscript.}

\data{All data that support the findings of this study are included within the article and its supplementary material. The three original AI-generated videos analysed in this study are provided as supplementary material (Videos 1--3). The extracted kinematic data and the corresponding analytical fits are presented in the article.}

\bibliographystyle{unsrt}
\bibliography{references}

\end{document}